\documentclass[a4paper,11pt]{article}
\usepackage{jinstpub}
\usepackage{lineno}

\title{\boldmath Reconstruction of the Impact Parameter in Nucleus-Nucleus Collisions at the MPD Experiment}

\author[1]{D.M. Idrisov\note{Corresponding author.}}
\author[1]{F.F. Guber}
\author[1]{N.M. Karpushkin}
\author[1,2]{P.E. Parfenov}

\affiliation[1]{Institute for Nuclear Research of the Russian Academy of Sciences,\\
Prospekt 60-letiya Oktyabrya 7a, Moscow, 117312, Russia}

\affiliation[2]{Joint Institute for Nuclear Research,\\
Joliot-Curie 6, Dubna, 141980, Russia}

\emailAdd{idrisov.dim@mail.ru}

\abstract{%
Event classification by centrality is one of the key tasks of the MPD (Multi-Purpose Detector) experiment at the NICA collider. Centrality characterizes the initial geometry of heavy-ion collisions through the correlation of measured observables with the impact parameter. Typically, charged-particle multiplicity serves as the observable of choice. However, this approach can introduce autocorrelation in net-proton multiplicity fluctuation studies. In this work, we propose a novel approach based on the combined use of signals from the forward hadron calorimeter (FHCal) and the electromagnetic calorimeter (ECal). This method is expected to suppress autocorrelation in the study of proton multiplicity fluctuations, while the combination of data from both detectors will enable unambiguous centrality event classification across the full centrality range.
}


\begin{document}
\maketitle
\flushbottom

\section{Introduction}
\label{sec:intro}

The search for signatures of deconfinement, a first-order phase transition, and the critical point in strongly interacting nuclear matter (QCD) forms the basis of the energy scan program for nuclear collisions in the center-of-mass energy range \(\sqrt{s_{NN}} = 4\)--11~GeV \cite{abgaryan2022, kekelidze2018, kapishin2020}. Investigating the QCD phase diagram at high baryon densities is a key scientific objective of the BM@N (Baryonic Matter at Nuclotron) and MPD (Multi-Purpose Detector) experiments at the NICA complex at JINR, Dubna \cite{kapishin2020}.

The use of the MPD detector in fixed-target mode (MPD-FXT) for relativistic heavy-ion collisions at $\sqrt{s_{NN}}$ = 2.3-3.5 GeV is currently under investigation~\cite{parfenov2025}. The MPD-FXT mode will enable not only a direct cross-checks of the data between the BM@N and MPD experiment at identical energies, but extend the acceptance coverage in comparison with the BM@N.

Observables sensitive to the equation of state of dense baryonic matter in heavy-ion collisions are strongly dependent on the initial collision geometry \cite{broniowski2002, alice2013, lacey2013, miller2007}. Centrality is a key parameter to assess the initial geometry through the correlation between measured observables and the impact parameter. Typically \cite{miller2007, loizides2015, d_enterria_progress_2021}, the Glauber model is used to determine centrality from charged-particle multiplicity measured in the central rapidity region by the time projection chamber (TPC). However, applying this approach at low energies presents several challenges.

First, using multiplicity alone can introduce autocorrelation when studying fluctuations, for example, in net-proton multiplicity measurements \cite{luo_volume_2013, na612025}. Second, a correct description of the interaction requires accounting for Coulomb scattering on the target nucleus \cite{mehndiratta2017}, as well as the increasing role of baryon stopping effects and energy conservation constraints,  led to the development of modified Glauber model versions \cite{simak2025}. The standard Glauber model approach reproduces charged-particle multiplicity with good accuracy down to approximately 3~GeV \cite{abdallah_measurements_2022}. However, the HADES experiment at an energy of 2.4~GeV (1.23A ~GeV) required a novel phenomenological parametrization, possibly indicating a change in collision dynamics in this energy regime \cite{hades2017}. Furthermore, the low and discrete multiplicities of produced particles at low energies complicate the optimization of centrality classes aimed at reducing volume fluctuations, which is a critical aspect in this energy regime \cite{luo_volume_2013}.

We  propose an alternative approach uses the energy measured by the electromagnetic calorimeter (ECal) and the energy of spectator fragments from the forward hadron calorimeter (FHCal). This approach provides independent information on the collision geometry and suppresses autocorrelation in net-proton fluctuation studies. To fully eliminate autocorrelation, the ECal signal can be selected such that its acceptance region does not overlap with the phase space of protons used in the fluctuation analysis, thereby completely excluding their contribution to centrality determination. An additional advantage of using calorimeters is their shorter timing window compared to tracking systems, which significantly reduces the contribution from pile-up effects at high beam intensities.

In this paper, we investigate centrality determination methods based on the direct reconstruction approach \cite{das2018} in comparison with  proposed new two-dimensional centrality determination method \citep{idrisov2025twodimensionalbayesianapproachcentrality} in the MPD-FXT configuration. The study is performed using model data from the DCM–QGSM–SMM \cite{baznat2020} generator for Xe+W collisions at a kinetic energy of \(E_{\text{kin}} = 2.5A\)~GeV. A realistic detector response is simulated with the GEANT4 toolkit \cite{allison2016} within the MPDROOT software framework \cite{rogachevsky2021}.

\section{MPD Experiment with a Fixed Target}

\begin{figure}[htbp]
\centering
\includegraphics[width=0.8\textwidth]{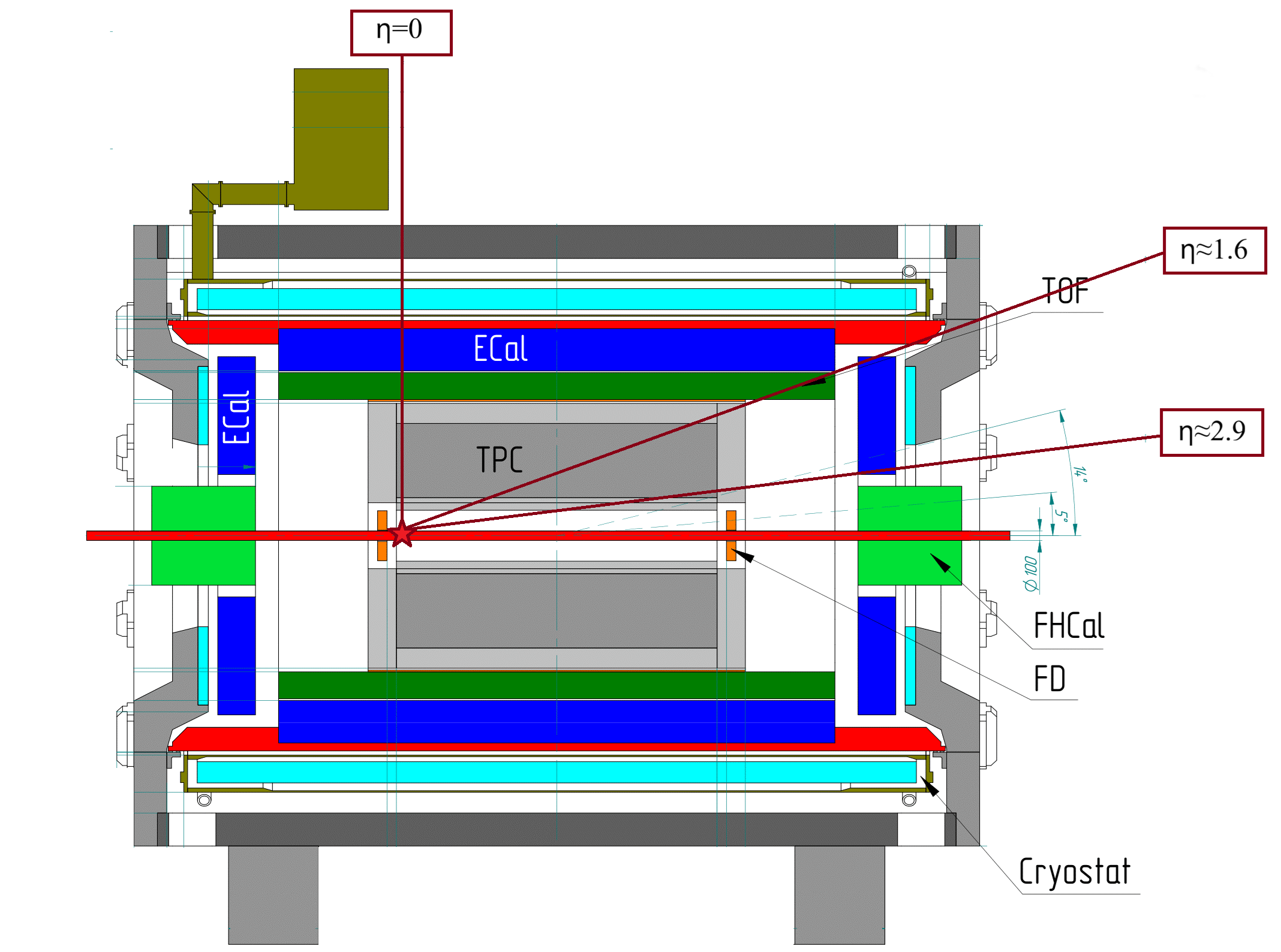}
\caption{Schematic view of the MPD-FXT setup \cite{parfenov2025}.}
\label{fig:setup}
\end{figure}

To implement the fixed-target research program at the MPD experiment, a thin wire target with a diameter of 50–100 \(\mu\)m is to be installed inside the NICA collider vacuum pipe. The target will be offset from the beam pipe center by approximately 1 cm and placed near the end of the central MPD detector. A schematic view of the experimental setup is shown in Fig.~\ref{fig:setup}. This configuration enables efficient detection of particles produced in nucleus-nucleus collisions.

The MPD detector in fixed-target mode incorporates the same main subsystems as in the collider mode \cite{abgaryan2022}: the time projection chamber (TPC), time-of-flight system (TOF), electromagnetic calorimeter (ECal), forward hadron calorimeter (FHCal), and the fast forward detector (FFD). Charged particles are identified via ionization energy loss (\(dE/dx\)) in the TPC gas and via the squared mass (\(m^2\)) derived from the time-of-flight measured by TOF and the momentum measured by TPC.

The forward hadron calorimeter FHCal measures the energy of spectator fragments — nucleons that did not participate in inelastic interactions. However, due to the central hole accommodating the vacuum pipe, the FHCal signal is not a monotonic function of the impact parameter: peripheral collisions with a low number of spectators can yield a signal comparable to that from central collisions. Resolving this ambiguity requires additional information, which is provided by the central electromagnetic calorimeter ECal. The ECal is designed for photon and electron detection with high energy resolution. Its signal is sensitive to the multiplicity of produced particles and decreases monotonically with increasing impact parameter. Combining data from FHCal and ECal enables unambiguous event classification across the entire centrality range.

\section{Methods for Centrality Determination}
To validate the two-dimensional centrality determination method proposed in this work, we compare its performance with the one-dimensional direct reconstruction approach based on charged-particle multiplicity. This approach, grounded in Bayes' theorem and the gamma-distribution description of multiplicity fluctuations, provides an independent way to reconstruct the impact parameter without relying on the Glauber model. The direct reconstruction method has been successfully applied to experimental data from ALICE, ATLAS, CMS, and LHCb \cite{rogly2018}, as well as in the low-energy regime of the INDRA experiment, where the classical Glauber-based approach becomes unreliable \cite{frankland_model_2021}.

\subsection{One-Dimensional Direct Reconstruction Method}

The direct reconstruction method demonstrates that recovering the impact parameter \(b\) from the measured multiplicity of charged particles \(N_{ch}\) constitutes a typical inverse problem solvable via Bayes' theorem \cite{rogly2018, das2018}. The core element is the fluctuation kernel, which models the fluctuations in the multiplicity of produced particles \(P(N_{ch}|b)\) at a fixed impact parameter \(b\). Studies \cite{rogly2018, parfenov2021} have shown that multiplicity fluctuations can be accurately described by a gamma distribution:

\begin{figure}[htbp]
\centering
\includegraphics[width=0.8\textwidth]{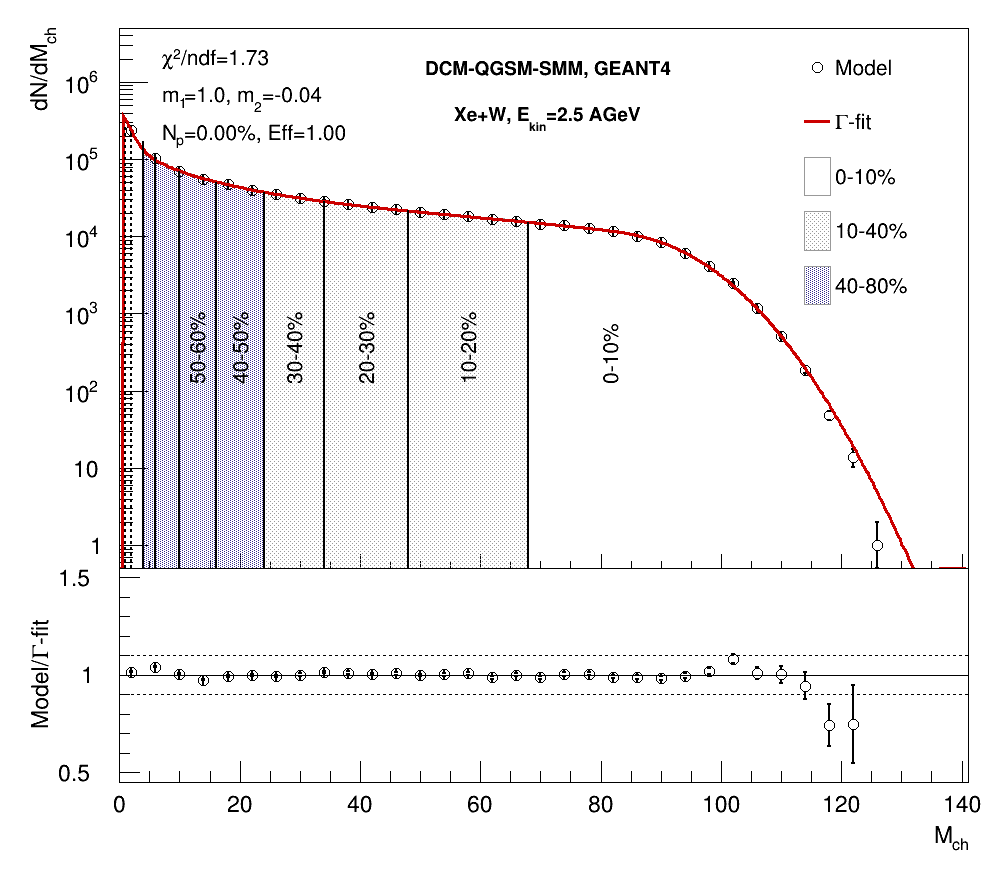}
\caption{Results of fitting the produced-particle multiplicity using the direct reconstruction method. The red line denotes the fit function, while open symbols correspond to data from the DCM-QGSM-SMM model for the Xe+W reaction at a beam energy of 2.5A GeV. Vertical lines indicate centrality classes.}
\label{fig:fit1d}
\end{figure}

\begin{equation}
P(N_{ch}|b) = \frac{1}{\Gamma(k)\theta^k} N_{ch}^{k-1} e^{-N_{ch}/\theta},
\label{eq:gamma}
\end{equation}

where \(\Gamma(k)\) is the gamma function. The parameters \(k\) and \(\theta\) are defined as \(k = \langle N_{ch} \rangle^2 / \sigma_{N_{ch}}^2\) and \(\theta = \sigma_{N_{ch}}^2 / \langle N_{ch} \rangle\).

The mean and variance of the observable can be derived from fully reconstructed model data under the following assumptions:

\begin{equation}
\langle N_{ch}(b) \rangle = \alpha_N \langle N_{ch}^{\text{MC}}(b) \rangle,\quad \sigma_{N_{ch}}^2(b) = \alpha_N^2 \sigma_{N_{ch}^{\text{MC}}}^2(b) + \alpha_N \beta_N \langle N_{ch}^{\text{MC}}(b) \rangle,
\label{eq:scaling}
\end{equation}

where \(\alpha_N, \beta_N\) are fit parameters accounting for the discrepancy between experimental data and simulation. The observed distribution \(P(N_{ch})\) is obtained from \(P(N_{ch}|b)\) as follows:

\begin{equation}
P(N_{ch}) = \int_0^1 P(N_{ch}|c_b) dc_b,
\label{eq:convolution}
\end{equation}

where \(c_b\) denotes centrality defined via the impact parameter. The results of applying the direct reconstruction method are shown in Fig.~\ref{fig:fit1d}.

\subsection{Two-Dimensional Direct Reconstruction Method}
\begin{figure}[htbp]
\centering
\includegraphics[width=\textwidth]{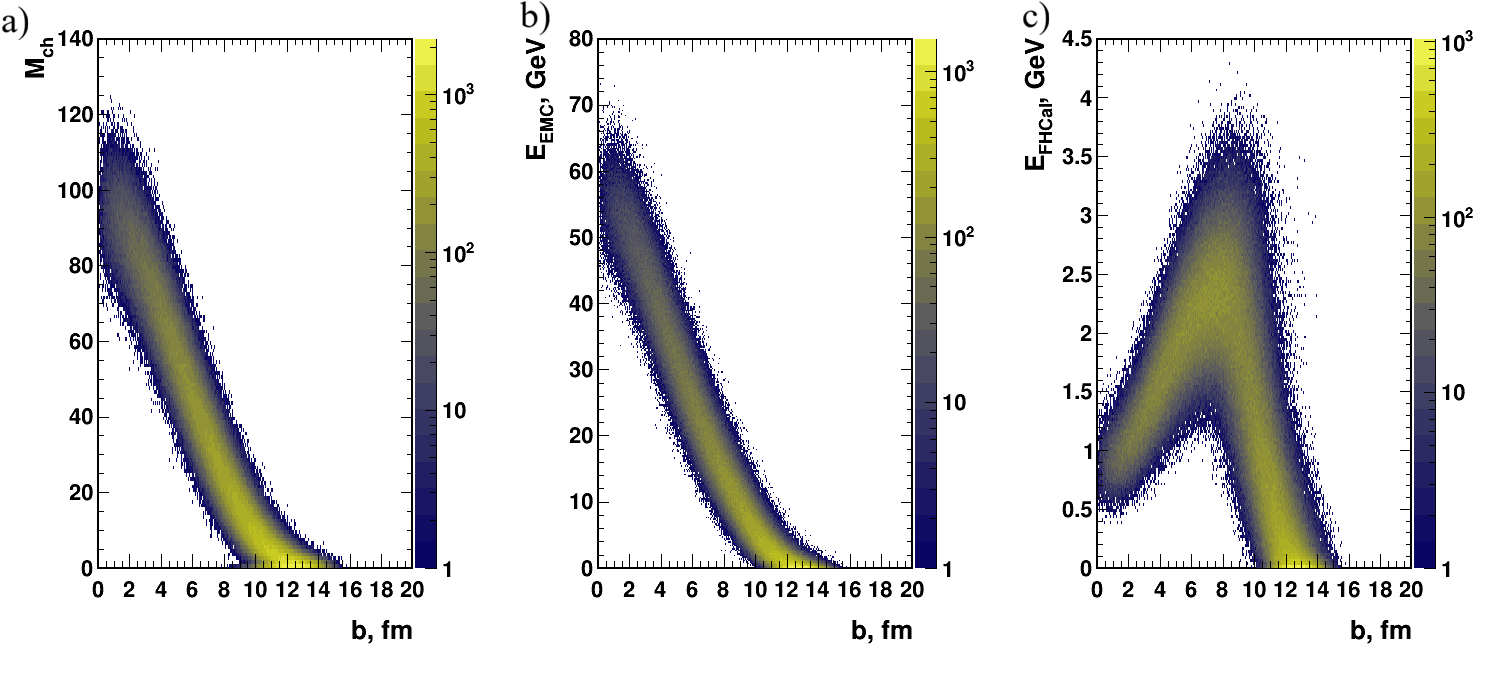}
\caption{(a) Correlation of charged-particle multiplicity with the impact parameter in the DCM-QGSM-SMM model \cite{baznat2020} for the Xe+W reaction at a beam energy of 2.5\(A\) GeV. (b) and (c) show correlations of energy measured by the electromagnetic calorimeter ECal and the forward hadron calorimeter FHCal with the impact parameter, respectively.}
\label{fig:correlations}
\end{figure}

Centrality determination using the FHCal and the ECal is a crucial task, as it provides an independent method for centrality estimation. Fig.~\ref{fig:correlations} displays correlations of various observables with the impact parameter within the DCM-QGSM-SMM model \cite{baznat2020} for the Xe+W reaction at a beam energy of 2.5\(A\) GeV.

The obtained data indicate that, due to the absence of a central module in the FHCal calorimeter, the correlation between its energy and the impact parameter is non-monotonic, as opposed to the ECal case. Therefore, these two observables are necessary to establish an unambiguous mapping of centrality classes in both central and peripheral regions. To address this, a direct reconstruction method for the two-dimensional case was developed, jointly analyzing signals from both calorimeters: the spectator energy from FHCal and the energy from ECal, which is sensitive to particles produced in the central rapidity region.
\begin{figure}[htbp]
\centering
\includegraphics[width=\textwidth]{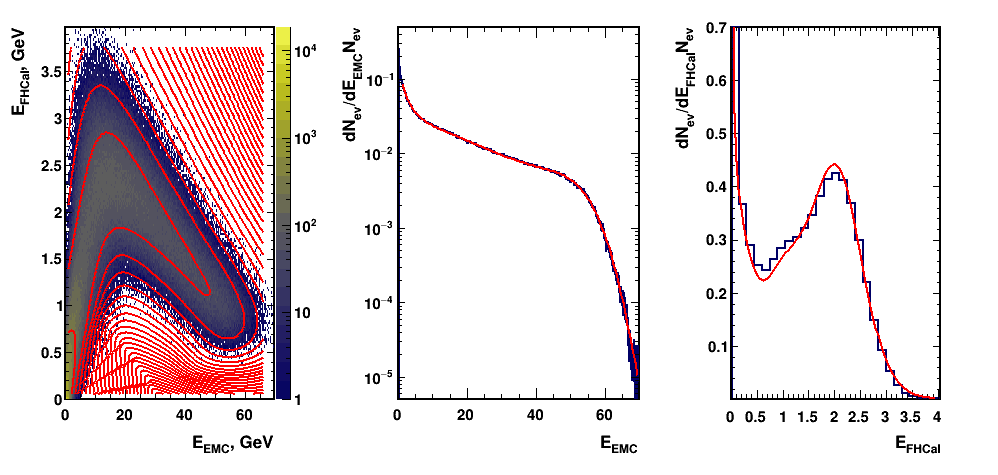}
\caption{(a) Two-dimensional distribution of energies from the FHCal and ECal detectors in the DCM-QGSM-SMM model for the Xe+W reaction at a beam energy of 2.5\(A\) GeV. Contour lines of the fit function are shown in red. (b) The ECal energy distribution is shown by the blue line; the projection of the fit function onto the ECal axis is depicted in red. (c) The energy distribution in the forward hadron calorimeter FHCal and the fit function are shown by blue and red lines, respectively.}
\label{fig:2dfit}
\end{figure}
The observed probability density distribution of the FHCal and ECal energies \(P(E_{\text{FHCal}}, E_{\text{ECal}})\) is related to the probability density at a fixed impact parameter \(P(E_{\text{FHCal}}, E_{\text{ECal}}|b)\) by:

\begin{equation}
P(E_{\text{FHCal}}, E_{\text{ECal}}) = \int_0^1 P(E_{\text{FHCal}}, E_{\text{ECal}}|c_b) dc_b.
\label{eq:2dconv}
\end{equation}

The relationship between the quantities \(X_1, X_2\) and \(E_{\text{FHCal}}, E_{\text{ECal}}\) is defined via a rotation matrix with an angle \(\phi\), chosen such that \(\text{Cov}(X_1, X_2) = 0\).

The total integral \(P(E_{\text{FHCal}}, E_{\text{ECal}})\) can be employed to fit the two-dimensional distribution of FHCal and ECal energies by assuming that the mean values and variances of the observables are proportional to those derived from fully reconstructed model data:

\begin{figure}[htbp]
\centering
\includegraphics[width=0.8\textwidth]{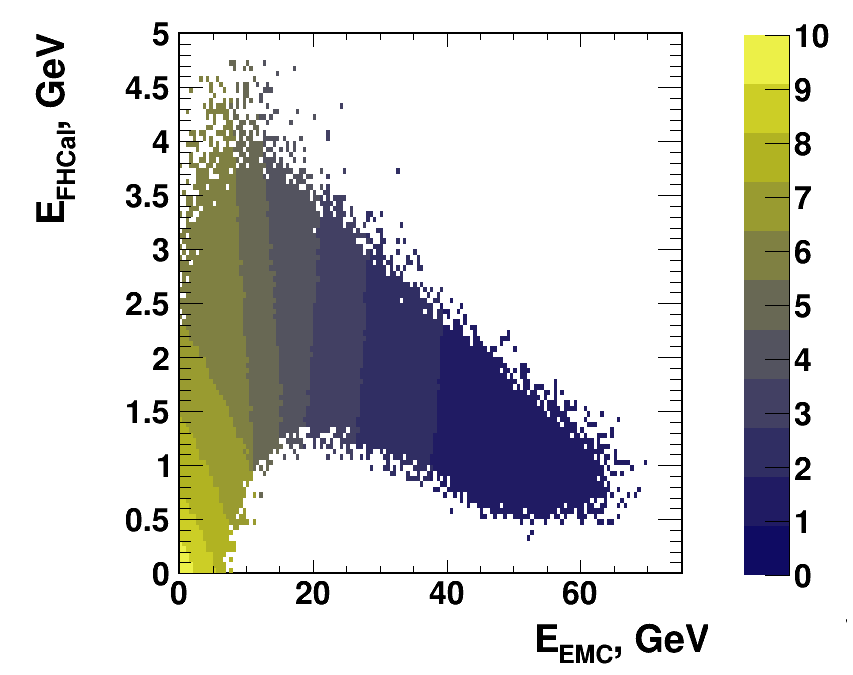}
\caption{Centrality classes for the energy distribution from the FHCal and ECal detectors in the DCM-QGSM-SMM model for the Xe+W reaction at a beam energy of 2.5\(A\) GeV.}
\label{fig:clusters}
\end{figure}

\begin{equation}
\begin{aligned}
\langle E_{\text{FHCal}}(b) \rangle &= \alpha_E \langle E_{\text{FHCal}}^{\text{MC}}(b)\rangle,\quad \langle E_{\text{ECal}}(b) \rangle = \alpha_N \langle E_{\text{ECal}}^{\text{MC}}(b)\rangle,\\
\sigma_{E_{\text{FHCal}}}^2(b) &= \alpha_E^{2}\sigma_{E_{\text{FHCal}}^{\text{MC}}}^2(b)+\alpha_E\beta_E\langle E_{\text{FHCal}}^{\text{MC}}(b) \rangle,\\
\sigma_{E_{\text{ECal}}}^2(b) &= \alpha_N^{2}\sigma_{E_{\text{ECal}}^{\text{MC}}}^2(b)+\alpha_N\beta_N\langle E_{\text{ECal}}^{\text{MC}}(b) \rangle,
\end{aligned}
\label{eq:2dscaling}
\end{equation}

where \(\alpha_E, \alpha_N, \beta_E, \beta_N\) are fit parameters.

The results of fitting the two-dimensional energy distribution from the FHCal and ECal detectors in the DCM-QGSM-SMM model for the Xe+W reaction at a beam energy of 2.5\(A\) GeV are shown in Fig.~\ref{fig:2dfit}. In Fig.~\ref{fig:2dfit}a, the red lines represent the contour lines of the fit function from Eq.~\ref{eq:2dconv}. Figs.~\ref{fig:2dfit}b and~\ref{fig:2dfit}c show the projections onto the ECal and FHCal axes, respectively. The obtained results demonstrate that the fit function reproduces the energy distribution in FHCal and ECal and can be used for centrality determination in nucleus-nucleus collisions.

Following the fitting procedure, the resulting two-dimensional distribution was divided into 10 centrality classes using the k-means-constrained method \cite{bradley2000}. This method minimizes the intra-cluster distance while also controlling the fraction of events within each cluster, thereby allowing the required event percentage to be correctly accounted for when defining centrality classes. The centrality classes obtained via the k-means-constrained method are depicted in Fig.~\ref{fig:clusters}.
\begin{figure}[htbp]
\centering
\includegraphics[width=0.55\textwidth]{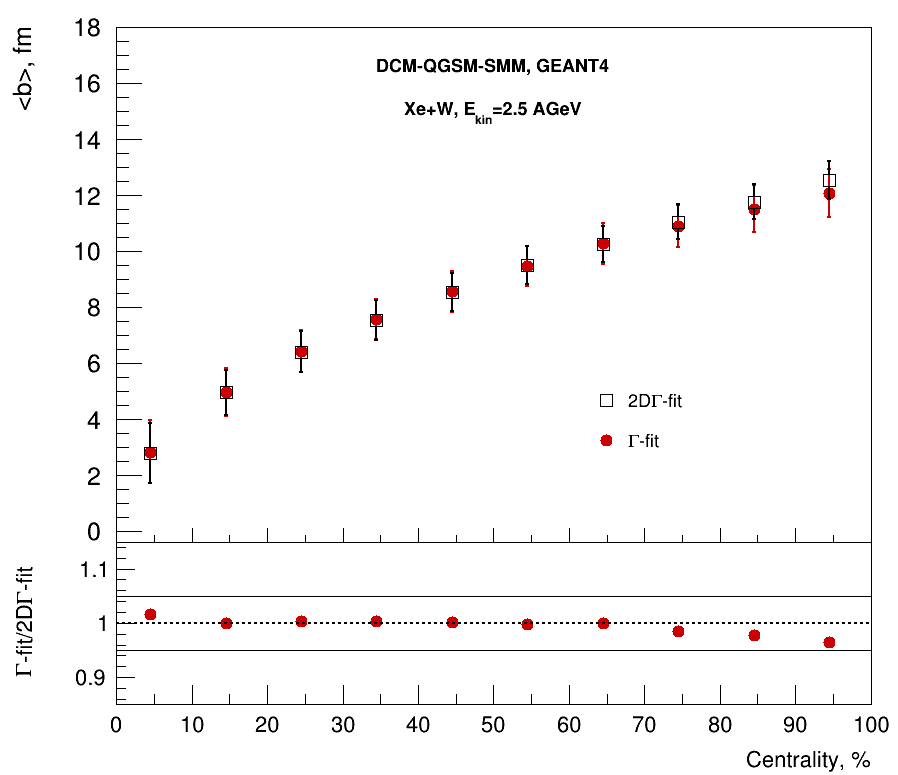}
\caption{Dependence of the mean impact parameter on centrality. Filled circles correspond to results obtained using the one-dimensional direct reconstruction method based on charged-particle multiplicity. Open squares represent results from the two-dimensional approach.}
\label{fig:comparison}
\end{figure}
Fig.~\ref{fig:comparison} presents a comparison of the mean impact parameter as a function of centrality derived from the two centrality determination methods. The obtained results are consistent within 5\%, confirming the efficacy of the two-dimensional direct reconstruction method. The proposed method can be utilized for studying proton multiplicity fluctuations and for other investigations where accounting for autocorrelation with the multiplicity of produced particles is important.

\clearpage

\section{Conclusion}

In this work, an extension of the direct reconstruction method to the two-dimensional case is proposed. It is shown that signals from both calorimeters -- the spectator energy from FHCal and the energy from ECal -- can be used for centrality classification of interaction events. The performance of the methods was validated using simulated data generated with DCM-QGSM-SMM for the Xe+W reaction at a beam energy of 2.5\(A\) GeV. It was shown that the results obtained with the two-dimensional method are consistent with the one-dimensional approach based on the multiplicity of produced particles, establishing it as a promising tool for studying proton multiplicity fluctuations in the MPD experiment. Using electromagnetic and hadron calorimeters for centrality determination will enable an independent assessment of the collision geometry and can reduce the impact of pile-up effects.

\acknowledgments

The authors express their gratitude to the MPD collaboration for fruitful discussions, valuable suggestions, and support.

\bibliographystyle{JHEP}
\bibliography{refs}

\end{document}